# Unveiling the Origin of the Basal-plane Antiferromagnetism in the $J_{\text{eff}}=1/2$ Mott Insulator Ba$_2$IrO$_4$: A Density Functional and Model Hamiltonian Study

Y. S. Hou, H. J. Xiang[*], and X. G. Gong[*]

Key Laboratory of Computational Physical Sciences (Ministry of Education), State Key Laboratory of Surface Physics, Collaborative Innovation Center of Advanced Microstructures, and Department of Physics, Fudan University, Shanghai 200433, P. R. China

Email: hxiang@fudan.edu.cn, xggong@fudan.edu.cn

**Abstract**

Based on the density functional theory and our new model Hamiltonian, we have studied the basal-plane antiferromagnetism in the novel $J_{\text{eff}}=1/2$ Mott insulator Ba$_2$IrO$_4$. By comparing the magnetic properties of the bulk Ba$_2$IrO$_4$ with those of the single-layer Ba$_2$IrO$_4$, we demonstrate unambiguously that the basal-plane antiferromagnetism is caused by the intralyer magnetic interactions rather than by the previously proposed interlayer ones. In order to reveal the origin of the basal-plane antiferromagnetism, we propose a new model Hamiltonian by adding the single ion anisotropy and pseudo-quadrupole interactions into the general bilinear pseudo-spin Hamiltonian. The obtained magnetic interaction parameters indicate that the single ion anisotropy and pseudo-quadrupole interactions are unexpectedly strong. Systematical Monte Carlo simulations demonstrate that the basal-plane antiferromagnetism is caused by the isotropic Heisenberg, bond-dependent Kitaev and pseudo-quadrupole interactions. Our results show for the first time that the single ion anisotropy and pseudo-quadrupole interaction can play significant roles in establishing the exotic magnetism in the $J_{\text{eff}}=1/2$ Mott insulator.

# I. INTRODUCTION

Recently, iridium oxides have been extensively studied both experimentally and theoretically because of their novel $J_{\text{eff}}=1/2$ states [1, 2] caused by the strong relativistic spin-orbit coupling (SOC). Many exotic and emerging phenomena are proposed to be realized in such iridium oxides. For example, quantum spin Hall Effect has been predicted in $Na_2IrO_3$ [3, 4]. Topological Mott insulator [5, 6], Weyl semimetal and axion insulator [7] are theoretically shown to exist in the pyrochlore iridium oxides. Most excitingly, iridium oxides with the honeycomb lattice are greatly expected to realize the long-sought Kitaev model, which has an exactly solvable quantum spin liquid ground state [8-11]. Initially, some theoretical and experimental studies suggested that the prototypical iridium oxide $Sr_2IrO_4$ (SIO) could display the unconventional high-$T_C$ superconductivity [12-15]. However, the $IrO_6$ octahedrons in SIO are rotationally distorted, resulting in an appreciable Dzyaloshinsky-Moriya (DM) interaction [16]. Consequently, the canted ferromagnetic (FM) moment induced by the DM interaction in SIO is about one order of magnitude larger than that in the layered $La_2CuO_4$ [16, 17] and the large FM moment may easily break the spin-singlet Cooper pairs. As a result, the rotational distortion of the $IrO_6$ octahedra should not favor the realization of the high-$T_C$ superconductivity in SIO and other iridium oxides need to be explored.

Different from SIO which crystallizes in the space group $I4_1/acd$ [18] and has the bent Ir-O-Ir bonds in the tetragonal $xy$ plane, the square-lattice antiferromagnet $Ba_2IrO_4$ (BIO) crystallizes in the space group $I4/mmm$ and has the straight Ir-O-Ir bonds in the tetragonal $xy$ plane [19]. In other words, the inversion symmetry at the midpoint of the adjacent and antiferromagnetically correlated $Ir^{4+}$ ions is preserved in BIO but it is broken in SIO. Therefore, the DM interaction automatically vanishes and no canted FM moments exist in BIO. Furthermore, the straight Ir-O-Ir bonds in BIO should give a large charge transfer integral, which may be favorable for metallization through carrier doping [20]. As a result, BIO is a closer analogue to the layered $La_2CuO_4$ than SIO. It is therefore expected that BIO is more suitable for realizing the high-$T_C$ superconductivity than SIO.

However, there remains a fundamental and important issue regarding the basal-plane antiferromagnetism in BIO to be clarified. Experimentally, it is found that the magnetic moments of $Ir^{4+}$ ions are antiferromagnetically correlated and point along the [110] direction [19-22].

Katukuri *et al.* theoretically investigated [23] how the [110]-pointing antiferromagnetism is established. Based on the general bilinear pseudo-spin Hamiltonian, they proposed that the experimentally observed basal-plane antiferromagnetism could be accounted for by including the additional interlayer magnetic interactions and the associated order-by-disorder quantum-mechanical effects [23]. However, the interlayer magnetic interactions in BIO may be too weak to be relevant to the experimentally observed basal-plane antiferromagnetism, because the shortest distance between the interlayer $Ir^{4+}$ ions is as far as about 7.3 Å [19] and the interlayer $Ir^{4+}$ ions are separated by two BaO layers. Therefore, it will be of great importance to reinvestigate the origin of the experimentally observed basal-plane antiferromagnetism in BIO.

In this paper, we put forward a new mechanism to explain the experimentally observed basal-plane antiferromagnetism in BIO, based on the density functional theory (DFT), our new model Hamiltonian, and Monte Carlo (MC) simulations. Comparison of the magnetic properties, especially the basal-plane magneto-crystalline anisotropy energy (MAE), between the bulk and single-layer BIO clearly demonstrates the experimentally observed basal-plane antiferromagnetism is caused by the intralyer magnetic interactions rather than by the previously proposed interlayer ones [23]. Next, it is proved that the calculated basal-plane MAE cannot be explained by the general bilinear pseudo-spin Hamiltonian. In order to reveal of the origin of the experimentally observed basal-plane antiferromagnetism, we propose a new model Hamiltonian by adding the single ion anisotropy (SIA) and pseudo-quadrupole interaction into the general bilinear pseudo-spin Hamiltonian. By computing the magnetic interaction parameters in our new model Hamiltonian, we find that the SIA and pseudo-quadrupole interaction are unexpectedly strong in BIO. Based on the systematical MC simulations, we demonstrate the experimentally observed basal-plane antiferromagnetism can be explained by the isotropic Heisenberg, bond-dependent Kitaev and pseudo-quadrupole interactions. Our results show the SIA and pseudo-quadrupole interaction can play significant roles in establishing the exotic magnetism in the $J_{eff}$=1/2 Mott insulator.

**II. COMPUTATIONAL METHODS**

Our first-principles calculations are based on the DFT and performed within the generalized gradient approximation (GGA) according to the Perdew-Burke-Ernzerhof (PBE) parameterization as implemented in Vienna *Ab initio* Simulation Package (VASP) [24]. We use the projector-augmented wave method [25] with an energy cutoff of 500 eV and a gamma-centered 7x7x3 k-point mesh grid. Ion positions are relaxed towards equilibrium with the Hellmann-Feynman forces on each ion less than 0.01 eV/Å. To describe the electron correlation associated with the 5*d* electron of $Ir^{4+}$ ions, we use the rotationally invariant DFT+U scheme introduced by Liechtenstein *et al.* [26]. The on-site Coulomb energy $U = 2.5 \text{ eV}$ and the Hund coupling $J_H = 0.5 \text{ eV}$ [2, 27, 28] are applied to the Ir 5*d* electrons. Because Ir is a kind of heavy element and its relativistic SOC ($\lambda = 0.45 \text{eV}$) [29, 30] is rather strong, SOC is taken into consideration in calculating the basal-plane MAE and the magnetic interactions parameters. The efficient exchange MC method is employed for finding the magnetic ground state [31, 32].

## III. RESULT AND DISCUSSION

In this section, we first demonstrate that the basal-plane antiferromagnetism in BIO is caused by the intralayer magnetic interactions instead of the previously proposed interlayer ones [23]. We then prove the basal-plane MAE of BIO can be phenomenologically accounted for by adding the SIA and pseudo-quadrupole interactions into the general bilinear pseudo-spin Hamiltonian. Finally, we demonstrate the experimentally observed basal-plane antiferromagnetism of BIO can be explained by the isotropic Heisenberg, bond-dependent Kitaev and pseudo-quadrupole interactions.

### A. Density Functional Theory Study on the Basal-plane Antiferromagnetism in BIO

Bulk BIO is a kind of $J_{\text{eff}}$=1/2 Mott insulator and has the quasi-two-dimensional lattice structure and tetragonal symmetry [19]. It consists of the single-layer BIO perovskite and these single-layer BIO perovskites stack along the *z* axis, i.e., the [001] direction (see Fig. 1a). There are two single-layer BIO perovskites in the unit cell of bulk BIO. In the tetragonal *xy* plane, $Ir^{4+}$ ions are located at the center of the corner-sharing oxygen octahedrons which have a 7% tetragonal distortion along the *z* axis [19]. The $IrO_2$ square-lattice planes are separated by two successive BaO layers and the $Ir^{4+}$ ions in the neighboring $IrO_2$ square-lattice planes are spatially separated as far as about 7.3 Å [19] (see Fig. 1a). Recent experiments show the crystal field splitting associated with the tetragonal distortion of the IrO$_6$ octahedron is small so that BIO is a close realization of the spin-orbit Mott insulator with the $J_{eff} = 1/2$ ground state [33].

We first confirm the magnetic ground state of the bulk BIO to be the antiferromagnetic (AFM) state with the magnetic moment of the Ir$^{4+}$ ions pointing along the [110] direction, in consistent with the experimental observation [19]. Here we take into account three representative AFM states: the AFM state with the magnetic moment of the Ir$^{4+}$ ions pointing along the [001] direction (denoted as AFM-001), along the [100] direction (denoted as AFM-100) and along the [110] direction (denoted as AFM-110) (see Fig. 1b). Although the total energy of the AFM-100 state is lower by 1.7 meV/Ir than that of the AFM-001 state, it is slightly higher by 0.19 meV/Ir than that of the AFM-110 state. It is worth of noting that if the initial state is the FM state, it cannot be stabilized and the resulting state is the paramagnetic state and has a higher total energy than the AFM-001, AFM-100 and AFM-110 states.

Then we investigate in details the basal-plane MAE of the bulk BIO and find that the basal-plane MAE can be fitted to a formula which includes the four-order contribution. Since the AFM state cannot be accommodated by the unit cell of the bulk BIO which has only one Ir$^{4+}$ ion per IrO$_2$ square-lattice plane, we use the $\sqrt{2} \times \sqrt{2} \times 1$ supercell containing two Ir$^{4+}$ ions per IrO$_2$ square-lattice plane. In order to obtain the basal-plane MAE, we constrain the magnetic moments of the antiferromagnetically correlated Ir$^{4+}$ ions in the tetragonal $xy$ plane and rotate them by an angle $\phi$ with respect to the reference [100] direction (see the inset of Fig. 1c). Fig. 1c shows the DFT+U+SOC calculated total energy dependence on the rotated angle $\phi$. As is expected, the minimum of the total energy occurs at the $\phi = 45°$, i.e., at the [110] direction. Because of the tetragonal symmetry, MAE can be approximated in the form of [34]

$$E(\theta,\phi) = K_1 \sin^2 \theta + K_2 \sin^4 \theta + K_3 \sin^4 \theta \sin^2 \phi \cos^2 \phi + E_0' \qquad (1).$$

In Eq. (1), $\theta$ and $\phi$ are the azimuthal and polar angles with respect to the [001] direction ($z$ axis) and [100] direction ($x$ axis), respectively. Because the magnetic moments of the Ir$^{4+}$ ions of the considered AFM states lie in the tetragonal $xy$ plane, $\theta$ is equal to $\pi/2$ and the first and second terms in Eq. (1) become constant. Thus the MAE in Eq. (1) reduces to

$$E(\theta = \pi/2, \phi) = K_3 \sin^2 \phi \cos^2 \phi + E_0 \qquad (2).$$

Note that the constant $K_1$ and $K_2$ in Eq. (1) are absorbed into the constant $E_0$. Based on Eq. (2), the calculated MAE parameter $K_3$ is $2.8 \times 10^6 \, \text{erg}/\text{cm}^3$, one order of magnitude larger than that of the body-center cubic iron [35].

To make clear the effect of the interlayer magnetic interactions on the basal-plane antiferromagnetism, we investigate the magnetic properties of the single-layer BIO. The lattice structure of the single-layer BIO is shown in the Fig. 2a. It consists of one of the single-layer BIO perovskite and vacuum layers. Because the $Ir^{4+}$ ions in the $IrO_2$ square-lattice planes are isolated by the thick enough vacuum layers, the interlayer magnetic interactions existing in the bulk BIO vanish in the single-layer BIO. Note that although the single-layer BIO (space group *P*4/mmm) has the slightly lower symmetry than the bulk BIO (space group *I*4/mmm), it preserves the vital tetragonal symmetry. Surprisingly, we find the interlayer magnetic interactions have no significant effect on the basal-plane antiferromagnetism, because the single-layer BIO almost reproduces the magnetic properties of the bulk BIO. First of all, the magnetic ground state of the single-layer BIO is the AFM-110 state. Secondly, the AFM-100 state has the higher total energy than the AFM-110 state but has the lower total energy than the AFM-001 state. Most importantly, the single-layer BIO has almost the same basal-plane MAE as the bulk BIO. Using the same method applied to the calculation of the basal-plane MAE of the bulk BIO, we obtain the basal-plane MAE of the single-layer BIO and show it in the Fig. 2b. By comparing the Fig. 1c and Fig. 2b, one can get the profile of the basal-plane MAE of the single-layer BIO is almost the same as that of the bulk BIO. Based on Eq. (2), the calculated MAE constant $K_3$ of the single-layer BIO is $2.9 \times 10^6 \, \text{erg}/\text{cm}^3$, very close to that of the bulk BIO. Therefore, it is clearly demonstrated that the experimentally observed basal-plane antiferromagnetism should be caused by the intralayer magnetic interactions instead of the interlayer ones. It also implies that the interlayer magnetic interactions are negligible compared with the intralayer ones, which will be discussed below (see Table *I*). Note that our result is distinctly different from the previous one that interlayer magnetic interactions can have significant effect on the experimentally observed basal-plane antiferromagnetism [23].

## B. Model Hamiltonian Study on the Basal-plane Antiferromagnetism in BIO

Here we first recall the Hamiltonian employed by the previous studies on the iridium oxides [8, 23, 36, 37]. The $Ir^{4+}$ ions yields an effective $J_{eff}=1/2$ Kramers-doublet ground states due to the strong octahedral crystal field and SOC [8, 21, 33]. The Kramers-doublet ground states are the pseudo-spin states $|\uparrow>$ and $|\downarrow>$ (denoted as $\overline{S}$ in order to distinguish from the usual spin $S$) which are in the form of [8]:

$$|\uparrow> = \sin\alpha |xy,\uparrow> + \frac{\cos\alpha}{\sqrt{2}}\left(i|xz,\downarrow> + |yz,\downarrow>\right) \quad (3.1)$$

$$|\downarrow> = \sin\alpha |xy,\downarrow> + \frac{\cos\alpha}{\sqrt{2}}\left(i|xz,\uparrow> - |yz,\uparrow>\right) \quad (3.2).$$

In Eq. (3.1) and (3.2), the angle $\alpha$ parameterizes the relative strength of the tetragonal crystal-field splitting and $\tan(2\alpha) = 2\sqrt{2}\lambda/(\lambda-\Delta)$ ($\Delta$ and $\lambda$ are the tetragonal crystal-field splitting and SOC parameter, respectively). For a pair of pseudo-spins $\overline{S}_i$ and $\overline{S}_j$, the general bilinear pseudo-spin Hamiltonian can be cast in the form of [8, 23, 36, 37]

$$H_{ij} = J_{ij}^H \overline{S}_i \cdot \overline{S}_j + J_{ij}^K \overline{S}_i^\gamma \overline{S}_j^\gamma + D_{ij} \cdot \left(\overline{S}_i \times \overline{S}_j\right) + J_{ij}^{DP} \left(\overline{S}_i \cdot r_{ij}\right)\left(r_{ij} \cdot \overline{S}_j\right) \quad (4).$$

In Eq. (4), $r_{ij}$ is the unit vector along the $ij$ bond and $\gamma$ is anyone of the $x$, $y$ and $z$. The first term is the isotropic Heisenberg interaction. The second term is the celebrated bond-dependent Kitaev interaction (see the ref. [8] for the details). The tetragonal symmetry of BIO requires that only the $J_{ij}^K \overline{S}_i^z \overline{S}_j^z$ term may not vanish [8], when the $z$ axis of the global coordinate system is set to be perpendicular to the $IrO_2$ square-lattice plane (see the global coordinate system plotted in the Fig. 1a). The third term is the DM interactions but it vanishes in BIO. The last term is the pseudo-dipolar interaction. For convenience, hereafter, the magnitude of the pseudo-spin is absorbed into the magnetic interactions parameters $J_{ij}^H$, $J_{ij}^K$ and $J_{ij}^{DP}$, so the pseudo-spin

We show here that the general pseudo-spin Hamiltonian defined by Eq. (4) cannot explain the calculated basal-plane MAE of the bulk BIO. Because the interlayer magnetic interactions are demonstrated to be negligible compared to the intralayer ones, they are left out of consideration here. We consider an arbitrary AFM state with the magnetic moments of the $Ir^{4+}$ ions pointing along the direction $(\theta,\phi)$ in the $\sqrt{2}\times\sqrt{2}\times1$ supercell and denote such state as $\text{AFM}(\theta,\phi)$. Firstly, the isotropic Heisenberg interaction has the SO(3) symmetry so it makes no contribution to the MAE. Secondly, the contribution from the nearest neighboring (NN) and next nearest neighboring (NNN) Kitaev interactions to the MAE is (see Appendix A)

$$E^K(J^K,\theta,\phi) = -8J_1^K \cos^2\theta + 8J_2^K \cos^2\theta \qquad (5).$$

In Eq. (5), $J_1^K$ and $J_2^K$ are the NN and NNN bond-dependent Kitaev interactions parameters, respectively. Thirdly, the contributions from the NN and NNN pseudo-dipole interactions to the MAE are (see the Appendix A)

$$E^{DP}(J^{DP},\theta,\phi) = -4J_1^{DP} \sin^2\theta + 4J_2^{DP} \sin^2\theta \qquad (6).$$

In Eq. (6), $J_1^{DP}$ and $J_2^{DP}$ are the NN and NNN pseudo-dipole interactions parameters, respectively. Therefore, the contribution from the NN and NNN bond-dependent Kitaev and pseudo-dipole interactions to the MAE of an arbitrary $\text{AFM}(\theta,\phi)$ state is only dependent on the azimuthal angle $\theta$ and independent on the polar angle $\phi$. We can further demonstrate that even if the negligible bond-dependent Kitaev and pseudo-dipole interactions of the Ir-Ir pairs farther than the NNN Ir-Ir one are taken into consideration, their contribution to the MAE still only depends on the azimuthal angle $\theta$ (see the Eq. (A.3) in the Appendix A). In summary, one can conclude based on the general pseudo-spin Hamiltonian defined by Eq. (4) that the basal-plane AFM state ($\theta = \pi/2$) is isotropic, which contradicts with the experimentally observed [110]-pointing AFM state [21].

In this work, we propose a new model Hamiltonian by adding the SIA and pseudo-quadrupole interaction into the general pseudo-spin Hamiltonian defined by Eq. (4). Although it is commonly believed that magnetic systems made up of spin-1/2 transition-metal ions have no magnetic anisotropy arising from SOC [38], our recent work showed that most spin-1/2 transition-metal ions did have the SIA [39, 40]. Because the pseudo-spin $\overline{S}$ is an analogy to the spin-1/2 state, it is natural to generalize that the pseudo-spin $\overline{S}$ does have the SIA which is in the form of

$A^{SIA}\left(\overline{S}^{z}\right)^{2}$. Note that the non-zero SIA has been pointed out to be present in the $J_{\text{eff}}=1/2$ state of Na$_2$IrO$_3$ [41] and SIO [42] (See Appendix B for a detailed discussion on the SIA of the $J_{\text{eff}}=1/2$ state). For the same reason, we further generalize that the quadrupole-quadrupole coupling proposed by Van Vleck [41] can also exist in the pseudo-spin $\overline{S}$ systems:

$$H_{ij}^{QP} = J_{ij}^{QP}\left(\overline{S}_{i}\cdot r_{ij}\right)^{2}\left(r_{ij}\cdot \overline{S}_{j}\right)^{2} \qquad (7).$$

In Eq. (7), $J_{ij}^{QP}$ and $r_{ij}$ are the pseudo-quadrupole interaction parameter and the unit vector along the $ij$ bond, respectively.

Now let us demonstrate the calculated basal-plane MAE can be well accounted for on the basis of our new model Hamiltonian. For an arbitrary AFM$(\theta,\phi)$ state in the $\sqrt{2}\times\sqrt{2}\times 1$ supercell, the contributions from the NN and NNN pseudo-quadrupole interactions to the MAE is (see the Appendix A)

$$E^{QP}\left(J^{QP},\theta,\phi\right) = \sin^{4}\theta\left[4J_{1}^{QP}\left(1-2\sin^{2}\phi\cos^{2}\phi\right)+2J_{2}^{QP}\left(1+4\sin^{2}\phi\cos^{2}\phi\right)\right] \qquad (8).$$

In Eq. (8), $J_{1}^{QP}$ and $J_{2}^{QP}$ are the NN and NNN pseudo-quadrupole interaction parameters, respectively. Since the pseudo-quadrupole interaction is a kind of short-range interaction, it is reasonable that only the NN and NNN pseudo-quadrupole interactions are taken into account. Taking into consideration the isotropic Heisenberg, bond-dependent Kitaev, pseudo-dipole and pseudo-quadrupole interactions of the NN and NNN Ir-Ir pairs and the SIA, we obtain their contribution to the MAE of an arbitrary AFM$(\theta,\phi)$ state as follows:

$$E(\theta,\phi) = K_{1}\sin^{2}\theta + K_{2}\sin^{4}\theta + K_{3}\sin^{4}\theta\sin^{2}\phi\cos^{2}\phi + E_{0} \qquad (9).$$

The coefficients in Eq. (9) are

$$K_{1} = 8J_{1}^{K} - 8J_{2}^{K} - 4J_{1}^{DP} + 4J_{2}^{DP} - 4A^{SIA} \qquad (9.1),$$

$$K_{2} = 4J_{1}^{QP} + 2J_{2}^{QP} \qquad (9.2),$$

$$K_{3} = -8J_{1}^{QP} + 8J_{2}^{QP} \qquad (9.3).$$

When the magnetic moments of the Ir$^{4+}$ ions lie on the tetragonal $xy$ plane, i.e., $\theta = \pi/2$, Eq. (9) reduces to $K_{3}\sin^{2}\phi\cos^{2}\phi + C$, which is the same as Eq. (2) and indicates that the basal-plane

AFM state is anisotropic instead of isotropic. Furthermore, if the coefficient $K_3$ is negative, the energy of the basal-plane AFM state has its minimum when the magnetic moments of the $Ir^{4+}$ ions point along the [110] direction. So the experimentally observed basal-plane antiferromagnetism can be phenomenologically explained by adding the pseudo-quadrupole interactions and the SIA into the generalized pseudo-spin Hamiltonian defined by Eq. (4).

Therefore the model Hamiltonian which well describes the magnetic behavior of BIO should be in the form of

$$H = \sum_{<i,j>\in xy}\left[ J_{ij}^H \overline{S}_i\cdot\overline{S}_j + J_{ij}^K \overline{S}_i^z \overline{S}_j^z + J_{ij}^{DP}\left(\overline{S}_i\cdot r_{ij}\right)\left(r_{ij}\cdot\overline{S}_j\right) + J_{ij}^{QP}\left(\overline{S}_i\cdot r_{ij}\right)^2\left(r_{ij}\cdot\overline{S}_j\right)^2 \right] +$$
$$\sum_{<<i,j>>\in xy}\left[ J_{ij}^H \overline{S}_i\cdot\overline{S}_j + J_{ij}^K \overline{S}_i^z \overline{S}_j^z + J_{ij}^{DP}\left(\overline{S}_i\cdot r_{ij}\right)\left(r_{ij}\cdot\overline{S}_j\right) + J_{ij}^{QP}\left(\overline{S}_i\cdot r_{ij}\right)^2\left(r_{ij}\cdot\overline{S}_j\right)^2 \right] +$$
$$\sum_i A_i^{SIA}\left(\overline{S}_i^z\right)^2 \qquad (10).$$

In Eq. (10), the first and second terms describe the magnetic interactions of the NN and NNN Ir-Ir pairs in the tetragonal *xy* plane, respectively. The isotropic Heisenberg, bond-dependent Kitaev, pseudo-dipole and pseudo-quadrupole interactions are all taken into consideration for them. The last term is the SIA generalized in our present work. Because the third nearest neighboring (NNNN) Ir-Ir pairs in the tetragonal *xy* plane and the interlayer Ir-Ir pairs are as far as 8.1 Å and 7.3 Å, respectively, their Heisenberg interactions should be rather weak. This is confirmed by our calculations (see Table *I*). Compared to the isotropic Heisenberg interaction, their bond-dependent Kitaev, pseudo-dipole and pseudo-quadrupole interactions should be much weaker and thereby left out of consideration hereafter.

The considered magnetic interactions and SIA parameters are obtained based on the DFT+U+SOC calculations and listed in the Table *I*. It is shown that the dominant magnetic interactions are the NN AFM and NNN FM Heisenberg interactions. So it is undoubted that the magnetic ground state should be AFM. Note that our calculated NN AFM Heisenberg interaction is well consistent with the previously calculated one [23, 43]. Besides, there are several important points being worth of noting. Firstly, it is beyond expectation that the NN bond-dependent Kitaev interaction has a much smaller magnitude than the NNN bond-dependent Kitaev interaction. Actually, this is reasonable and understandable. The NN bond-dependent Kitaev interaction is realized through the 180-degree Ir-O-Ir bond so it should be rather weak [8], as the calculated one.

However, it has been shown that there exists a large Kitaev interaction in slightly distorted 90-degree Ir-O-Ir bond of $Na_2IrO_3$ [8, 44]. Therefore it is expected that the four distorted Ir-O-Ir bonds (116.5-degree) facilitate and result in the unexpectedly large NNN Kitaev interaction (see the inset of Fig. 3a). Secondly, the pseudo-quadrupole interaction and the SIA are both strong, which is a manifestation of the strong SOC in iridium oxides. Finally, the NNNN and interlayer Heisenberg interactions are very weak, as expected. Note that the former and the latter are smaller by one order and by two orders than the NN Heisenberg interaction, respectively. An important reason for such weak interlayer Heisenberg interaction is that two BaO layers embed into the neighboring $IrO_2$ planes and prevent the electron from hopping between the interlayer $Ir^{4+}$ ions. Because the NNNN and interlayer Heisenberg interactions are much weaker than the NN and NNN Heisenberg interactions, they will be left out of consideration in the below discussion.

In order to reveal how the bond-dependent Kitaev, pseudo-dipole and pseudo-quadrupole interactions and the SIA affect the antiferromagnetism of BIO, we investigate the dependence of the energy differences $\Delta E1=E(AFM-001)-E(AFM-110)$ and $\Delta E2=E(AFM-100)-E(AFM-110)$ on them. Based on Eq. (9), the energy difference $\Delta E1$ and $\Delta E2$ are calculated by linearly increasing the NN and NNN bond-dependent Kitaev interaction parameters from the vanishing ones ($J^K/J_0^K=0$) to the DFT+U+SOC calculated ones ($J^K/J_0^K=1$) while keeping other magnetic interaction parameters unchanged. The similar method is applied to calculate the dependence of $\Delta E1$ and $\Delta E2$ on the pseudo-dipole and pseudo-quadrupole interactions and the SIA parameters. Fig. 3a shows the energy difference $\Delta E1$ gets larger and larger as the bond-dependent Kitaev interaction becomes stronger and stronger. This indicates the bond-dependent Kitaev interaction does not favor the magnetic moments of the $Ir^{4+}$ ions to point along the $z$ axis. In the Fig. 3b, 3c and 3d, the energy difference $\Delta E1$ gets smaller and smaller as the corresponding pseudo-dipole, pseudo-quadrupole interactions and the SIA become stronger and stronger. This indicates the pseudo-dipole, pseudo-quadrupole interactions, and the SIA favor the magnetic moments of the $Ir^{4+}$ ions to point along the $z$ axis. Because the energy difference $\Delta E2$ plotted in the Fig. 3a, 3b and 3d has nothing to do with the corresponding bond-dependent Kitaev, pseudo-dipole interactions and SIA, one can conclude that they are not relevant to the fact that the magnetic moments of the $Ir^{4+}$ ions point along the [110]-direction. However, the energy difference $\Delta E2$ plotted in the Fig. 3c indicates the AFM-110 and AFM-100 states are degenerate if the pseudo-quadrupole interaction is vanished and this degeneracy can be broken by the non-vanished pseudo-quadrupole interaction. Furthermore, the energy difference ΔE2 gets larger and larger as the pseudo-quadrupole interaction becomes stronger and stronger. So the pseudo-quadrupole interaction drives the magnetic moments of the $Ir^{4+}$ ions to point along the [110]-direction.

Finally, we show the experimentally observed AFM-110 state is caused by the isotropic Heisenberg, bond-dependent Kitaev and pseudo-quadrupole interactions, based on the systematical MC simulations (see Table *II*). Because the isotropic Heisenberg interaction is the dominant magnetic interaction, it is always included in the MC simulations. However, the bond-dependent Kitaev, pseudo-dipole and pseudo-quadrupole interactions and the SIA are optionally included in the MC simulations. By comparing the Case 1 and 2 listed in the Table *II*, one can obtain that, for the antiferromagnetically correlated $Ir^{4+}$ ions, the bond-dependent Kitaev interaction makes their magnetic moments lie at the tetragonal *xy* plane with the continuous degeneracy. However, Case 7 clearly shows the pseudo-quadrupole interaction can break the above-mentioned continuous degeneracy and the resulting magnetic structure is just the experimentally observed AFM-110 state. By carefully investigating the Table *II*, we can obtain that: (1) The magnetic ground state is the basal-plane AFM state with the magnetic moments of the $Ir^{4+}$ ions lying in the tetragonal *xy* plane as long as the Heisenberg and Kitaev interactions are taken into account; (2) The magnetic ground state is the experimentally observed AFM-110 state as long as the isotropic Heisenberg, bond-dependent Kitaev and pseudo-quadrupole interactions are taken into account; (3) Although the pseudo-dipole, pseudo-quadrupole interactions, and the SIA favor the magnetic moments of the $Ir^{4+}$ ions to point along the *z* axis (see the Case 3, 4 and 5 in the Table *II*), they cannot destroy the basal-plane AFM state established by the isotropic Heisenberg and bond-dependent Kitaev interactions. To sum up, it is clearly shown that the experimentally observed AFM-110 state can be explained by the isotropic Heisenberg, bond-dependent Kitaev and pseudo-quadrupole interactions.

## IV. SUMMARY

In summary, we have studied the basal-plane antiferromagnetism in the novel $J_{eff}=1/2$ Mott insulator $Ba_2IrO_4$. We show the basal-plane antiferromagnetism is caused by the intralayer magnetic interactions. We propose a new model Hamiltonian by adding the single ion anisotropy and pseudo-quadrupole interaction into the general bilinear pseudo-spin Hamiltonian to reveal the origin of the basal-plane antiferromagnetism. We find that the single ion anisotropy and pseudo-quadrupole interaction are unexpectedly strong. Based on the systematical MC simulations, we reveal the experimentally observed basal-plane antiferromagnetism can be explained by the isotropic Heisenberg, bond-dependent Kitaev and pseudo-quadrupole interactions. Our study suggests the SIA and pseudo-quadrupole interaction can play significant roles in establishing the exotic magnetism in many other $J_{\text{eff}}$=1/2 Mott insulators.

**ACKNOWLEDGEMENT**

This paper was partially supported by the National Natural Science Foundation of China, the Special Funds for Major State Basic Research, the Foundation for the Author of National Excellent Doctoral Dissertation of China, the Program for Professor of Special Appointment at Shanghai Institutions of Higher Learning, and the Research Program of Shanghai Municipality, the Ministry of Education, and Fok Ying Tung Education Foundation. We are grateful to Dr. Jihui Yang for his valuable discussion.

**Figure**

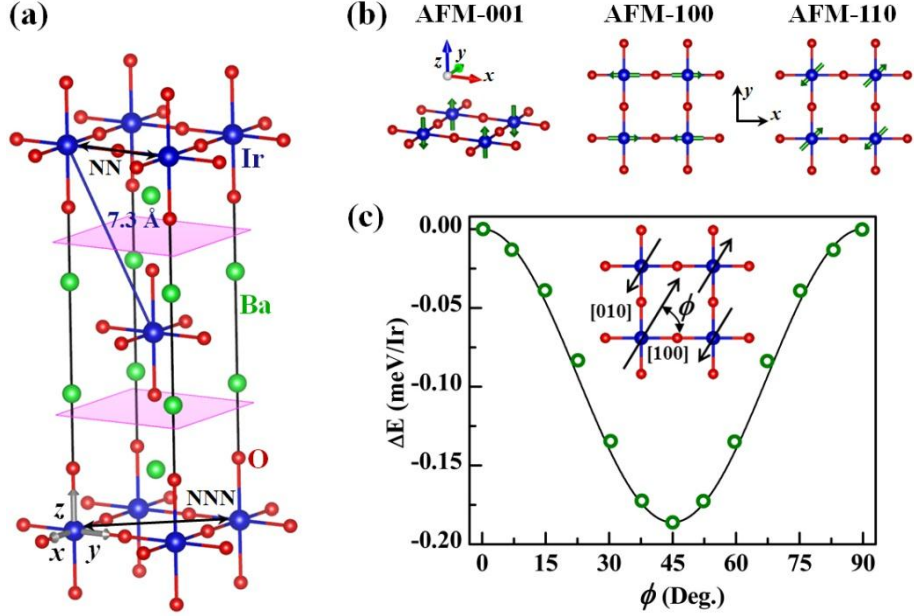

Fig. 1 (Color online) (a) Lattice structure of the bulk $Ba_2IrO_4$. Green, blue and red spheres represent Ba, Ir and O atoms, respectively. The NN and NNN Ir-Ir pairs are connected by black double-arrow solid lines. The shortest distance (7.3 Å) between the two $Ir^{4+}$ ions in the neighboring $IrO_2$ square-lattice planes is shown by the solid blue line. The structure unit, i.e., the single-layer $Ba_2IrO_4$ perovskite, is sandwiched by the two pink rectangles. The global *xyz* coordinate system is shown by the gray arrows. Note that the [100], [010] and [001] directions correspond to *x*, *y*, *z* axes, respectively. (b) The magnetic structures of the representative AFM-001, AFM-100 and AFM-110 states. The magnetic moments of the $Ir^{4+}$ ions are represented by green arrows. (c) The dependence of the total energy of the basal-plane AFM state on the polar angle $\phi$. The open green circles show the total energy obtained by the DFT+U+SOC calculations. The solid line shows the fitted total energy based on the MAE defined by Eq. (2). The inset shows the definition of the polar angle $\phi$ of the basal-plane AFM state with respect to the [100] direction.

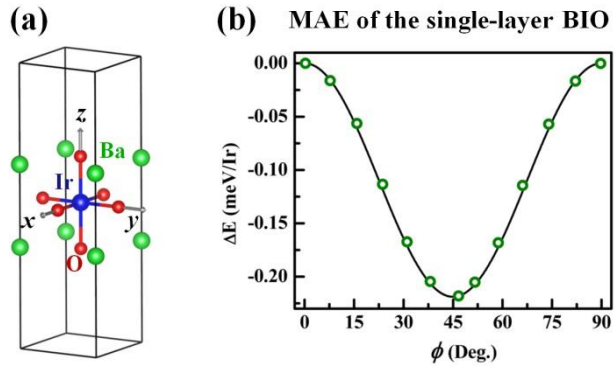

Fig. 2 (Color online) (a) Lattice structure of the single-layer Ba$_2$IrO$_4$. Ba, Ir and O atoms are represented by the green, blue and red spheres, respectively. The global *xyz* coordinate system is shown by the gray arrows. (b) The dependence of the total energy of the basal-plane AFM state on the polar angle $\phi$. The open green circles show the total energy obtained by the DFT+U+SOC calculations. The solid line shows the fitted total energy based on the MAE defined by Eq. (2).

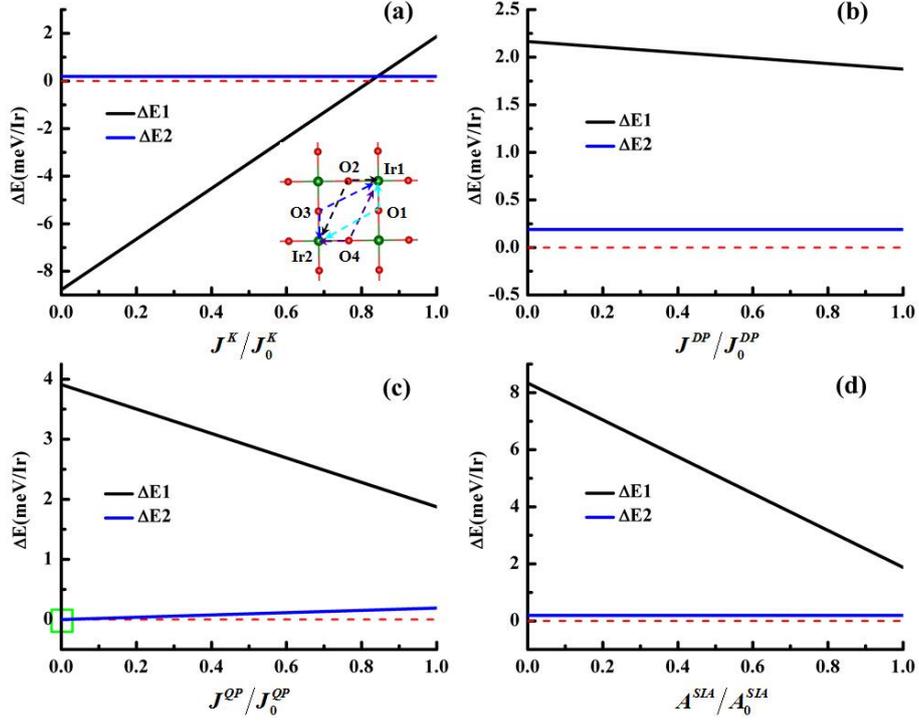

Fig. 3 (Color online) The dependence of the energy differences ΔE1=E(AFM-001)-E(AFM-110) (black line) and ΔE2=E(AFM-100)-E(AFM-110) (blue line) on the bond-dependent Kitaev interaction (a), pseudo-dipole interaction (b), pseudo-quadrupole interaction (c) and SIA (d). $J_0^K$, $J_0^{DP}$, $J_0^{QP}$ and $A_0^{SIA}$ are the DFT+U+SOC calculated bond-dependent Kitaev, pseudo-dipole and pseudo-quadrupole interactions and the SIA parameters, respectively. The red lines in (a), (b), (c) and (d) are for eye guidance. In (a), the inset shows the four 116.5-degree Ir-O-Ir bonds of the specified NNN Ir1-Ir2 pair. The four Ir-O-Ir bonds are depicted by the dashed black, blue, purple and cyan lines. In (c), the point ($J^{QP}/J_0^{QP}=0$, ΔE2=0) is highlighted by the green square.

Table *I*. The considered magnetic interaction and SIA parameters are obtained by means of the DFT+U+SOC calculations and given in units of meV. The calculated SIA parameter is $A^{SIA} = -6.46$ meV. As for the NNNN and interlayer Ir-Ir pairs, only the isotropic Heisenberg interactions are calculated.

| Ir-Ir pairs | $J^H$ | $J^K$ | $J^{DP}$ | $J^{QP}$ |
|---|---|---|---|---|
| NN | 36.17 | -0.75 | -0.14 | 1.61 |
| NNN | -11.46 | 4.58 | 0.15 | 1.23 |
| NNNN | 1.92 | -- | -- | -- |
| Interlayer | -0.14 | -- | -- | -- |

Table *II*. Magnetic ground states obtained from systematical MC simulations in which different magnetic interactions are included. Plus sign (+) means the corresponding magnetic interaction is included in the MC simulations while the minus sign (−) means the corresponding magnetic interaction is not included in the MC simulations. The two-fold degeneracy of the AFM-001 state results from the time reversal invariance and the eight-fold degeneracy of the AFM-110 state results from the time reversal invariance and the four-fold rotation symmetry around the *z* axis.

| Case | $J^H$ | $J^K$ | $J^{DP}$ | $J^{QP}$ | $A^{SIA}$ | Ground State | Direction | Degeneracy |
|---|---|---|---|---|---|---|---|---|
| 1  | + | − | − | − | − | AFM | any | SO(3) |
| 2  | + | + | − | − | − | AFM | *xy* plane | continuous |
| 3  | + | − | + | − | − | AFM | z | 2 |
| 4  | + | − | − | + | − | AFM | z | 2 |
| 5  | + | − | − | − | + | AFM | z | 2 |
| 6  | + | + | + | − | − | AFM | *xy* plane | continuous |
| 7  | + | + | − | + | − | AFM | 110 | 8 |
| 8  | + | + | − | − | + | AFM | *xy* plane | continuous |
| 9  | + | − | + | + | − | AFM | z | 2 |
| 10 | + | − | + | − | + | AFM | z | 2 |
| 11 | + | − | − | + | + | AFM | z | 2 |
| 12 | + | + | + | + | − | AFM | 110 | 8 |
| 13 | + | + | + | − | + | AFM | *xy* plane | continuous |
| 14 | + | + | − | + | + | AFM | 110 | 8 |
| 15 | + | − | + | + | + | AFM | z | 2 |
| 16 | + | + | + | + | + | AFM | 110 | 8 |

## Appendix A: The dependence of MAE on the Kitaev, pseudo-dipole and pseudo-quadrupole Interactions

In this Appendix, we provide some details of the study on the MAE of the bond-dependent Kitaev, pseudo-dipole and pseudo-quadruple interactions. Here we consider an arbitrary $\text{AFM}(\theta,\phi)$ state (see the definition in the main text) in the $\sqrt{2}\times\sqrt{2}\times 1$ supercell. For a given Ir-Ir pair denoted as pair $i$, the magnetic moments of this Ir-Ir pair are correlated either antiferromagnetically or ferromagnetically. For example, the magnetic moments of the NN Ir-Ir pair are antiferromagnetically correlated while the magnetic moments of the NNN Ir-Ir pair are ferromagnetically correlated. The polar angle of a given Ir-Ir pair $i$ with respect to the $x$ axis is denoted as $\varphi_i$. For example, the polar angle $\varphi_{NN}$ of the NN Ir-Ir pair is $\pi/4$. As a result of the four-fold rotation symmetry around the $z$ axis, there exist other three symmetrically equal Ir-Ir pairs with respect to a given Ir-Ir pair $i$. The polar angles of these four symmetrically equal Ir-Ir pairs are $\varphi_i$, $\varphi_i+\pi/2$, $\varphi_i+\pi$ and $\varphi_i+3\pi/2$, respectively.

The contribution to the MAE from the Kitaev interaction of a given Ir-Ir pair $i$ is

$$E_i^K(\theta,\phi) = 8C_i J_i^K \cos^2\theta \qquad (\text{A.1.0}).$$

In Eq. (A.1.0), $J_i^K$ is the Kitaev interaction parameter and $C_i$ is $-1$ $(+1)$ if the given Ir-Ir pair is antiferromagnetically (ferromagnetically) correlated. Especially, the contributions from the NN and NNN Ir-Ir pairs are

$$E_1^K(\theta,\phi) = -8J_1^K \cos^2\theta \qquad (\text{A.1.1})$$

and

$$E_2^K(\theta,\phi) = 8J_2^K \cos^2\theta \qquad (\text{A.1.2}),$$

respectively. Summing over all the Ir-Ir pairs farther than the NNN Ir-Ir pairs, we can obtain

$$E^K(\theta,\phi) = \sum_i E_i^K(\theta,\phi) = \cos^2\theta \sum_i 8C_i J_i^K = J^K \cos^2\theta \qquad (\text{A.1}).$$

As for the contribution from the pseudo-dipole interaction of a given Ir-Ir pair $i$ to the MAE, it is

$$\begin{aligned}E_i^{DP}(\theta,\phi) &= 2C_i J_i^{DP} \sin^2\theta \left[\cos^2(\phi-\varphi_i) + \cos^2(\phi-\varphi_i-\pi/2)\right] + \\ &\quad 2C_i J_i^{DP} \sin^2\theta \left[\cos^2(\phi-\varphi_i-\pi) + \cos^2(\phi-\varphi_i-3\pi/2)\right] \\ &= 4C_i J_i^{DP} \sin^2\theta \qquad (\text{A.2.0}).\end{aligned}$$

Especially, the contributions from the NN and NNN Ir-Ir pairs are

$$E_1^{DP}(\theta,\phi) = -4J_1^{DP} \sin^2\theta \qquad (A.2.1)$$

and

$$E_2^{DP}(\theta,\phi) = 4J_2^{DP} \sin^2\theta \qquad (A.2.2),$$

respectively. Therefore, the contribution from the pseudo-dipole interaction of the Ir-Ir pairs farther than the NNN Ir-Ir pairs is

$$E^{DP}(\theta,\phi) = \sum_i E_i^{DP}(\theta,\phi) = \sin^2\theta \sum_i 4C_i J_i^{PD} = J^{DP}\sin^2\theta \qquad (A.2).$$

Based on the Eq. (A.1) and (A.2), we obtain the contribution from the Kitaev and pseudo-dipole interactions of the Ir-Ir pairs farther than the NNN Ir-Ir pairs to the MAE is

$$E(J^K, J^{DP}, \theta, \phi) = E^K(\theta,\phi) + E^{DP}(\theta,\phi) = J^K \cos^2\theta + J^{DP}\sin^2\theta \qquad (A.3).$$

Eq. (A.3) clears show the MAE is only dependent on the azimuthal angle $\theta$ but independent on the polar angle $\phi$.

The contribution from the pseudo-quadrupole interaction of the NN Ir-Ir pair to the MAE is

$$\begin{aligned}E_1^{QP}(\theta,\phi) &= 2J_{NN}^{DP} \sin^4\theta \left[\cos^4\phi + \cos^4(\phi-\pi/2) + \cos^4(\phi-\pi) + \cos^4(\phi-3\pi/2)\right] \\ &= 4J_{NN}^{DP} \sin^4\theta \left[\cos^4\phi + \sin^4\phi\right] \\ &= 4J_{NN}^{DP} \sin^4\theta \left[1 - 2\sin^2\phi\cos^2\phi\right] \qquad (A.4.1).\end{aligned}$$

As for the contribution from the pseudo-quadrupole interaction of the NNN Ir-Ir pair to the MAE, it is

$$\begin{aligned}E_2^{QP}(\theta,\phi) &= 2J_{NNN}^{DP} \sin^4\theta \left[\cos^4\left(\phi-\frac{\pi}{4}\right) + \cos^4\left(\phi-\frac{3\pi}{4}\right) + \cos^4\left(\phi-\frac{5\pi}{4}\right) + \cos^4\left(\phi-\frac{7\pi}{4}\right)\right] \\ &= 4J_{NNN}^{DP} \sin^4\theta \left[\cos^4(\phi-\pi/4) + \sin^4(\phi-\pi/4)\right] \\ &= 4J_{NNN}^{DP} \sin^4\theta \left[1 - 2\sin^2(\phi-\pi/4)\cos^2(\phi-\pi/4)\right] \\ &= 2J_{NNN}^{DP} \sin^4\theta \left[1 + 4\sin^2\phi\cos^2\phi\right] \qquad (A.4.2).\end{aligned}$$

**Appendix B: Single-Ion Anisotropy of the $J_{\text{eff}}=1/2$ State in the Tetragonal Crystal Field**

The single ion model Hamiltonian of the $Ir^{4+}$ ion in the cubic crystal field is read as

$$H_{SOC} = \lambda \mathbf{L}\cdot\mathbf{S} \quad (\lambda > 0) \quad\quad (B.1).$$

Because the energies of $e_g$ orbitals are far above those of the $t_{2g}$ orbitals, $H_{SOC}$ is restricted to the $t_{2g}$ subspace $\{|xz,\uparrow>, |yz,\uparrow>, |xy,\downarrow>, |xz,\downarrow>, |yz,\downarrow>, |xy,\uparrow>\}$ and its matrix presentation is

$$H_{SOC} = \begin{array}{c|cccccc} & xz\uparrow & yz\uparrow & xy\downarrow & xz\downarrow & yz\downarrow & xy\uparrow \\ \hline xz\uparrow & 0 & -i\frac{\lambda}{2} & i\frac{\lambda}{2} & 0 & 0 & 0 \\ yz\uparrow & i\frac{\lambda}{2} & 0 & -\frac{\lambda}{2} & 0 & 0 & 0 \\ xy\downarrow & -i\frac{\lambda}{2} & -\frac{\lambda}{2} & 0 & 0 & 0 & 0 \\ xz\downarrow & 0 & 0 & 0 & 0 & i\frac{\lambda}{2} & i\frac{\lambda}{2} \\ yz\downarrow & 0 & 0 & 0 & -i\frac{\lambda}{2} & 0 & \frac{\lambda}{2} \\ xy\uparrow & 0 & 0 & 0 & -i\frac{\lambda}{2} & \frac{\lambda}{2} & 0 \end{array}$$

Note that in the $t_{2g}$ subspace the effective total angular momentum operator $\mathbf{J}_{\text{eff}}$ is $\mathbf{J}_{\text{eff}} = \mathbf{S} - \mathbf{L}$ and the commutation relations are $\left[J_{\text{eff}}^{\alpha}, \mathbf{L}\cdot\mathbf{S}\right]=0 \ (\alpha=x,y,z)$ and $\left[\mathbf{J}_{\text{eff}}^2, \mathbf{L}\cdot\mathbf{S}\right]=0$. Therefore $J_{\text{eff}}^x$, $J_{\text{eff}}^y$, $J_{\text{eff}}^z$, and $\mathbf{J}_{\text{eff}}$ are all good quantum numbers and the six eigenstates can be labeled as

$$|1_C> = |J_{\text{eff}} = 3/2, J_{\text{eff}}^z = 3/2>,$$

$$|2_C> = |J_{\text{eff}} = 3/2, J_{\text{eff}}^z = 1/2>,$$

$$|3_C> = |J_{\text{eff}} = 3/2, J_{\text{eff}}^z = -1/2>,$$

$$|4_C> = |J_{\text{eff}} = 3/2, J_{\text{eff}}^z = -3/2>,$$

$$|5_C> = |J_{\text{eff}} = 1/2, J_{\text{eff}}^z = 1/2>,$$

$$|6_C> = |J_{\text{eff}} = 1/2, J_{\text{eff}}^z = -1/2>.$$

Their corresponding eigenvalues are $-\lambda/2$, $-\lambda/2$, $-\lambda/2$, $-\lambda/2$, $\lambda$ and $\lambda$. Because $Ir^{4+}$ ions have five 5d electrons, the ground state is a single hole residing on the $|J_{eff}=1/2, J_{eff}^z=1/2>$ or $|J_{eff}=1/2, J_{eff}^z=-1/2>$ states (Kramers degeneracy). Hereafter, we discuss the $J_{eff}$=1/2 states in the language of holes. Because operators $J_{eff}^x$, $J_{eff}^y$ and $J_{eff}^z$ commutate with the Hamiltonian $H_{SOC}$, $|J_{eff}=1/2, J_{eff}^x=1/2>$ and $|J_{eff}=1/2, J_{eff}^y=1/2>$ states are the linear combination of the $|J_{eff}=1/2, J_{eff}^z=1/2>$ and $|J_{eff}=1/2, J_{eff}^z=-1/2>$ states, namely,

$$|J_{eff}=\frac{1}{2}, J_{eff}^x=\frac{1}{2}>=\frac{1}{\sqrt{2}}\left(|J_{eff}=\frac{1}{2}, J_{eff}^z=\frac{1}{2}>+|J_{eff}=\frac{1}{2}, J_{eff}^z=-\frac{1}{2}>\right),$$

$$|J_{eff}=\frac{1}{2}, J_{eff}^y=\frac{1}{2}>=\frac{1}{\sqrt{2}}\left(|J_{eff}=\frac{1}{2}, J_{eff}^z=\frac{1}{2}>+i|J_{eff}=\frac{1}{2}, J_{eff}^z=-\frac{1}{2}>\right).$$

It can be verified that the energy expectations of the $|J_{eff}=1/2, J_{eff}^x=1/2>$, $|J_{eff}=1/2, J_{eff}^y=1/2>$ and $|J_{eff}=1/2, J_{eff}^z=1/2>$ states are equal, so the pure $J_{eff}$=1/2 states in the cubic crystal field have no single-ion anisotropy (SIA). Therefore the lack of SIA of the pure $J_{eff}$=1/2 states is protected by the cubic symmetry.

In the tetragonal crystal field, the Hamiltonian is read as

$$H = H_{SOC} + H_\Delta = \lambda \mathbf{L} \cdot \mathbf{S} + \Delta \times L_z^2 \qquad (B.2).$$

In this case, the operators $J_{eff}^x$, $J_{eff}^y$, and $J_{eff}^2$ do not commute with the Hamiltonian Eq. (B.2) because of $\left[J_{eff}^x, L_z^2\right] \neq 0$, $\left[J_{eff}^y, L_z^2\right] \neq 0$ and $\left[J_{eff}^2, L_z^2\right] \neq 0$, which indicates that the $J_{eff}$=1/2 states in the tetragonal crystal field are not pure. However, $J_{eff}^z$ commutates with the Hamiltonian Eq. (B.2) because of $\left[J_{eff}^z, L_z^2\right] = 0$. Therefore $J_{eff}^z$ is still a good quantum number in the presence of the tetragonal crystal field. Furthermore, the expectation $\langle J_{eff}^2 \rangle$ of the operator $J_{eff}^2$ is close to $\frac{3}{2} \times \left(\frac{3}{2}+1\right)$ and $\frac{1}{2} \times \left(\frac{1}{2}+1\right)$ (see figure A1). So the six eigenstates of the Hamiltonian Eq. (B.2) can be still labeled by $J_{eff}^2$ and $J_{eff}^z$, namely,

$$|1_T\rangle = \left|J_{eff} \approx \frac{3}{2}, J_{eff}^z = \frac{3}{2}\right\rangle,$$

$$|2_T\rangle = \left|J_{eff} \approx \frac{3}{2}, J_{eff}^z = \frac{1}{2}\right\rangle,$$

$$|3_T\rangle = \left|J_{eff} \approx \frac{3}{2}, J_{eff}^z = -\frac{1}{2}\right\rangle,$$

$$|4_T\rangle = \left|J_{eff} \approx \frac{3}{2}, J_{eff}^z = -\frac{3}{2}\right\rangle,$$

$$|5_T\rangle = \left|J_{eff} \approx \frac{1}{2}, J_{eff}^z = \frac{1}{2}\right\rangle,$$

$$|6_T\rangle = \left|J_{eff} \approx \frac{1}{2}, J_{eff}^z = -\frac{1}{2}\right\rangle.$$

The ground state is a single hole residing on the non-pure $J_{eff}=1/2$ states

$$|5_T\rangle = |J_{eff} \approx \frac{1}{2}, J_{eff}^z = \frac{1}{2}\rangle = \sin\alpha\,|xy,\uparrow\rangle + \frac{\cos\alpha}{\sqrt{2}}\left(i\,|xz,\downarrow\rangle + |yz,\downarrow\rangle\right) \quad (B.3.1)$$

or

$$|6_T\rangle = |J_{eff} \approx \frac{1}{2}, J_{eff}^z = -\frac{1}{2}\rangle = \sin\alpha\,|xy,\downarrow\rangle + \frac{\cos\alpha}{\sqrt{2}}\left(i\,|xz,\uparrow\rangle - |yz,\uparrow\rangle\right) \quad (B.3.2).$$

In Eq. (B.3.1) and (B.3.2) the angle $\alpha$ parameterizes the relative strength of the tetragonal crystal-field splitting and $\tan(2\alpha) = 2\sqrt{2}\lambda/(\lambda-\Delta)$ ($\Delta$ is the tetragonal crystal-field splitting). Because the operator $J_{eff}^x$ does not commutate with the Hamiltonian Eq. (B.2), the $|J_{eff} \approx 1/2, J_{eff}^x = 1/2\rangle$ state is not a linear combination of the $|J_{eff} \approx 1/2, J_{eff}^z = 1/2\rangle$ and $|J_{eff} \approx 1/2, J_{eff}^z = -1/2\rangle$ states. Actually, the desired $|J_{eff} \approx 1/2, J_{eff}^x = 1/2\rangle$ state is the linear combination of the non-pure $J_{eff}=3/2$ and $J_{eff}=1/2$ states, that is,

$$\left|J_{eff} \approx \frac{1}{2}, J_{eff}^x = \frac{1}{2}\right\rangle = C_1|1_T\rangle + C_2|2_T\rangle + C_3|3_T\rangle + C_4|4_T\rangle + C_5|5_T\rangle + C_6|6_T\rangle \quad (B.4).$$

In Eq. (B.4) the coefficients $C_i$ (i=1, 2, 3, 4, 5, 6) should satisfy such constraints:

$$C_1^\dagger C_1 + C_2^\dagger C_2 + C_3^\dagger C_3 + C_4^\dagger C_4 + C_5^\dagger C_5 + C_6^\dagger C_6 = 1,$$

$$\left\langle J_{eff}\approx\frac{1}{2},J_{eff}^{x}=\frac{1}{2}\left|J_{eff}^{x}\right|J_{eff}\approx\frac{1}{2},J_{eff}^{x}=\frac{1}{2}\right\rangle =\frac{1}{2},$$

$$\left\langle J_{eff}\approx\frac{1}{2},J_{eff}^{x}=\frac{1}{2}\left|J_{eff}^{2}\right|J_{eff}\approx\frac{1}{2},J_{eff}^{x}=\frac{1}{2}\right\rangle =\left\langle J_{eff}\approx\frac{1}{2},J_{eff}^{z}=\frac{1}{2}\left|J_{eff}^{2}\right|J_{eff}\approx\frac{1}{2},J_{eff}^{z}=\frac{1}{2}\right\rangle .$$

So it is natural to expect that the energy expectation of the $|J_{eff}\approx1/2,J_{eff}^{x}=1/2>$ state is different from that of the $|J_{eff}\approx1/2,J_{eff}^{z}=1/2>$ state. Whatever the tetragonal crystal field splitting $\Delta$ is, the ground state of the $Ir^{4+}$ ion is a single hole residing on the $|J_{eff}\approx1/2,J_{eff}^{z}=1/2>$ state. In order to produce the desired $|J_{eff}\approx1/2,J_{eff}^{x}=1/2>$ state, the single hole should have the possibility of residing on the lower-energy $|J_{eff}\approx3/2>$ state. For the hole, the lower-energy state is not favored. So the z axis is the easy axis of the $J_{eff}=1/2$ state of the $Ir^{4+}$ ion in the tetragonal crystal field, which is consistent our DFT calculated results. Therefore it is the symmetry lowering from the cubic to the tetragonal that makes the $J_{eff}=1/2$ states have the SIA.

In order to confirm the above results numerically, we introduce a new Hamiltonian $H^{new}$ by adding a constraint term $\alpha\cdot J_{eff}$ onto the Hamiltonian Eq. (B.2):

$$H^{new}=H_{SOC}+H_{\Delta}+H_{constraint}=\lambda L\cdot S+\Delta\times L_{z}^{2}+\alpha\cdot J_{eff} \qquad (B.5).$$

The constraint term is merely used to construct the desirable state. In Eq. (B.5), the parameter $\alpha$ is a real vector and in the form of $\alpha=(\alpha_{x},\alpha_{y},\alpha_{z})$. If the parameter $\alpha$ is set to be $\alpha=(\alpha_{x},0,0)$ ($\alpha_{x}>0$), the Eq. (B.5) will select out a specific hole state $|J_x>$ which approaches to the $|J_{eff}\approx1/2,J_{eff}^{x}=1/2>$ state. Likewise, the Eq. (B.5) will select out a specific hole state $|J_z>$ which approaches to the $|J_{eff}\approx1/2,J_{eff}^{z}=1/2>$ state if the parameter $\alpha$ is set to be $\alpha=(0,0,\alpha_{z})$ ($\alpha_{z}>0$). For the selected hole states $|J_x>$ and $|J_z>$, their energy expectation are calculated based on the formulas E($J_x$)=<$J_x|H_{SOC}+H_\Delta|J_x$> and E($J_z$)=<$J_z|H_{SOC}+H_\Delta|J_z$> such that the effect of the constraint is removed. Note that the denotation E($J_z=1/2$) in the Figure A2 is the energy of the hole eigenstate $|J_{eff}\approx1/2,J_{eff}^{z}=1/2>$ of the Eq. (B.2). Because the operator

$J_{eff}^z$ still commutes with the Hamiltonian Eq. (B.5) in the case of $\alpha = (0,0,\alpha_z)$, the selected hole state $|J_z\rangle$ is the hole eigenstate $|J_{eff} \approx 1/2, J_{eff}^z = 1/2\rangle$ of the Eq. (B.2), which is consistent with the vanishing energy difference $\Delta E_z = E(J_z) - E(J_z=1/2)$ and the constant $\langle J_z\rangle = 0.5$ (see the Figure B2). It is clearly shown in the Figure B2 that the selected holes state $|J_x\rangle$ has a higher energy than the hole eigenstate $|J_{eff} \approx 1/2, J_{eff}^z = 1/2\rangle$ no matter the tetragonal distortion is elongation ($\Delta>0$) or compression ($\Delta<0$) along the z axis, which indicates that the easy axis is the z axis. Therefore the non-pure $J_{eff}=1/2$ state in $Ba_2IrO_4$ with the tetragonal symmetry has SIA and its easy axis is the z axis, which is consistent with our DFT calculated results.

**Figures of Appendix B**

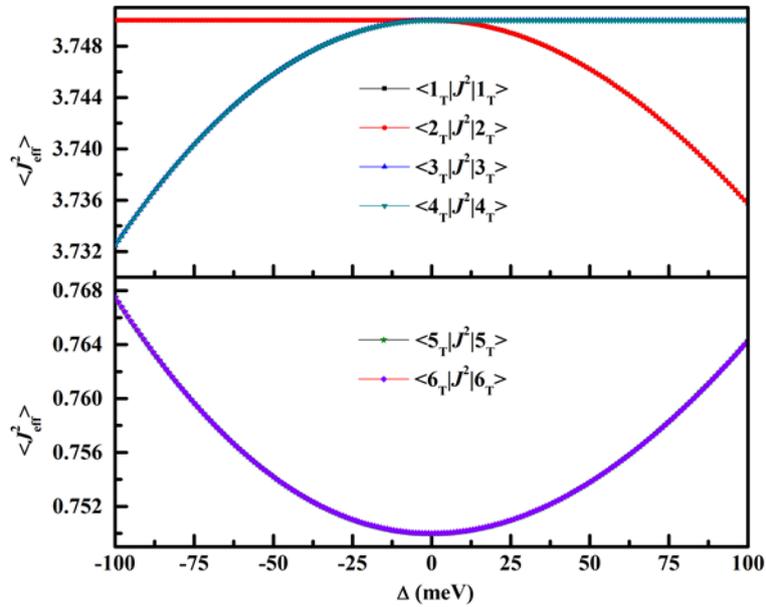

Figure B1. Expectation $\langle J_{eff}^2 \rangle$ of the six eigenstates labeled by $|i_T\rangle \, (i=1,2,3,4,5,6)$ of the Hamiltonian Eq. (B.2).

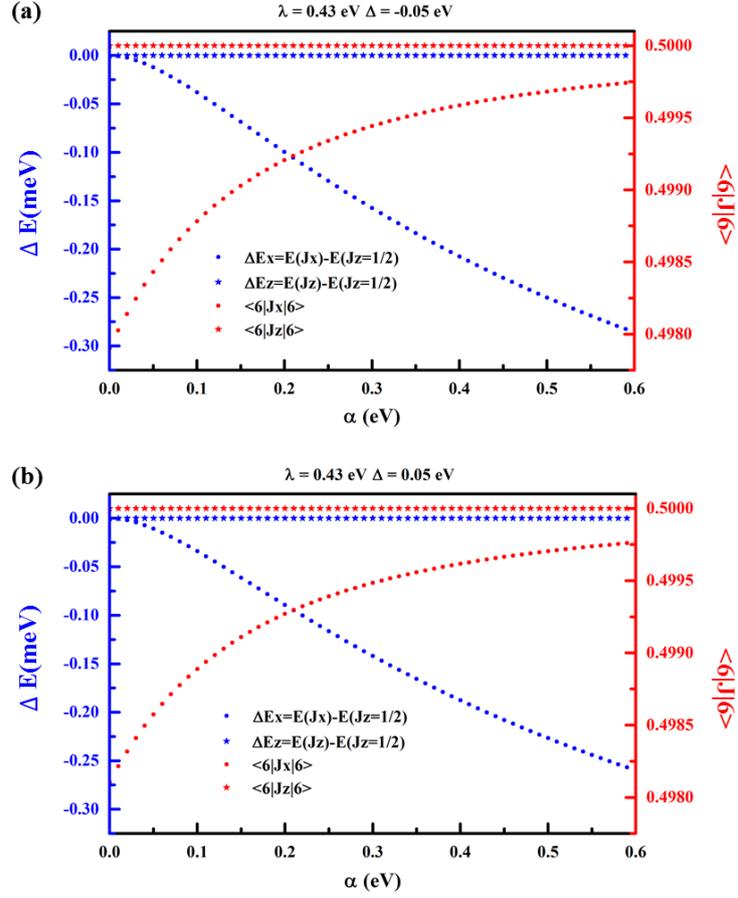

Figure B2. The dependence on the constraint parameter $\alpha$ of the expectation $<J_x>$ (read circle), $<J_z>$ (read star) and energy difference $\Delta E_x = E(J_x) - E(J_z=1/2)$ (blue circle), $\Delta E_z = E(J_z) - E(J_z=1/2)$ (blue star) of the selected hole states based on the Hamiltonian Eq. (B.5) in the tetragonal crystal field splitting $\Delta < 0$ (a) and $\Delta > 0$ (b). The selected hole state is denoted by $|6>$. Here we set the tetragonal crystal field splitting $\Delta$ to be 50 meV according to the Ref. [33].